\documentclass[reqno]{amsart}
\usepackage{epsf}
\usepackage{color}
\def\be{\begin{equation}}
\def\ee{\end{equation}}
\def\ba{\begin{array}}
\def\ea{\end{array}}
\def\bea{\begin{eqnarray}}
\def\eea{\end{eqnarray}}
\def\drm{{\mathrm d}}
\def\erm{{\mathrm e}}
\def\dps{\displaystyle}

\def\prt{\partial}

\def\la{\grbf^2 \,}

\def\gr{\grbf \,}

\def\gr{\mbox{\boldmath $\nabla$} }

\def\la{\gr^2}

\newcommand{\nc}{\newcommand}
\nc{\tcb}{\textcolor{blue}}
\nc{\tcr}{\textcolor{red}}
\nc{\tcg}{\textcolor{green}}

\nc{\qq}{\qquad\qquad}
\nc{\dis}{\displaystyle}
\nc{\ug}{\; = \;}
\nc{\Ebf}{\mbox{\boldmath $E$}}
\nc{\Bbf}{\mbox{\boldmath $B$}}
\nc{\abf}{\mbox{\boldmath $a$}}
\nc{\vbf}{\mbox{\boldmath $v$}}
\nc{\Fbf}{\mbox{\boldmath $F$}}
\nc{\rbf}{\mbox{\boldmath $r$}}
\nc{\Jbf}{\mbox{\boldmath $J$}}
\nc{\rd}{{\rm d}}
\nc{\dtau}{{\rd\tau}}
\nc{\dt}{{\rd t}}
\nc{\omf}{{\frac{\omega}{\omz}}}
\nc{\tz}{\tau_0}

\begin{document}

\vspace{1truecm}

\title{Majorana solutions to the two-electron problem}

\author{S. Esposito}
\address{{\it S. Esposito}: I.N.F.N. Sezione di
Napoli, Complesso Universitario di M. S. Angelo, Via Cinthia,
80126 Naples, Italy ({\rm Salvatore.Esposito@na.infn.it})}%

\author{A. Naddeo}
\address{{\it A. Naddeo}: Dipartimento di Fisica ``E.R. Caianiello'',
Universit\`a di Salerno, and CNISM Unit\`a di Ricerca di Salerno, Via Ponte don Melillo, 84084 Fisciano (Salerno), Italy
 ({\rm naddeo@sa.infn.it})}%


\begin{abstract}
A review of the known different methods and results devised to study the two-electron atom problem, appeared in the early years of quantum mechanics, is given, with particular reference to the calculations of the ground state energy of helium. This is supplemented by several, unpublished results obtained around the same years by Ettore Majorana, which results did not convey in his published papers on the argument, and thus remained unknown until now. Particularly interesting, even for current research in atomic and nuclear physics, is a general variant of the variational method, developed by Majorana in order to take directly into account, already in the trial wavefunction, the action of the full Hamiltonian operator of a given quantum system. Moreover, notable calculations specialized to the study of the two-electron problem show the introduction of the remarkable concept of an effective nuclear charge {\it different} for the two electrons (thus generalizing previous known results), and an application of the perturbative method, where the atomic number $Z$ was treated effectively as a {\it continuous} variable, contributions to the ground state energy of an atom with given $Z$ coming also from any other $Z$. Instead, contributions relevant mainly for pedagogical reasons count simple broad range estimates of the helium ionization potential, obtained by suitable choices for the wavefunction, as well as a simple alternative to Hylleraas' method, which led Majorana to first order calculations comparable in accuracy with well-known order 11 results derived, in turn, by Hylleraas.
\end{abstract}

\maketitle



\section{Introduction}

\noindent It is commonly believed that the early wide acceptance of quantum mechanics was triggered by the successful description of the simplest atomic system, i.e. the hydrogen atom, according to already known spectroscopic data. This is usually deduced from the fact that the analytic solution for the wavefunction of the (isolated) hydrogen atom indeed played a pivotal role in subsequent applications of the new quantum theory, during the early part of the XX century. However, historians of physics know quite well \cite{Mehra} that the definitive abandonment  of the old quantum theory by N. Bohr and A. Sommerfeld came through the thorough consideration of the helium atom problem, with particular reference to the specific prediction of the ground state energy of neutral helium (or other measured quantities related to it).

In 1913, following the discovery of the atomic nucleus \cite{nucleus}, Bohr succeeded \cite{Bohr1913} in explaining the energy levels of the hydrogen atom in terms of quantization of the action for the classical Kepler orbits. Numerous attempts to explain the ground state of helium bu quantizing different two-electron periodic orbits in a similar manner, however, failed. For example, Bohr firstly discussed a simple model where both electrons in the helium atom move along the same circular orbit and are located in the opposite ends of a diameter \cite{Bohr1913}. This followed the 1904 proof  by the Japanese H. Nagaoka \cite{Nagaoka} (obviously in the framework of classical mechanics) that such motion is mechanically stable (for sufficiently large attractive forces), and does have the lowest possible energy. In general, the attempts to quantize the helium atom in the Bohr-Sommerfeld theory \cite{Sommerfeld} were always based on the assumptions that the ground state is related to a single periodic orbit of the electron pair, and that the electrons move on symmetric orbits with equal radii for all times. Furthermore, it was as well assumed that orbits where the electrons hit the nucleus are excluded, and that the quantum number $n$ in the quantization condition is an integer. Such approach proved unsuccessful mainly because it failed to explain the diamagnetism of a two-electron atom in its lowest energy state and the value of the obtained energy was only in a rather poor agreement with experiments. 

As reported by Briggs \cite{Briggs}, in a letter to Sommerfeld of 1922, W. Heisenberg sketched a model where the electrons moved asymmetrically in opposite directions, and the quantum number associated to the quantization of the relative motion of the orbits under the influence of the mutual electron interaction was allowed to be {\it half-integer}. Interestingly enough, Heisenberg predicted a value for the measured ionization potential very close to that claimed by experiments of the time \cite{Sommerfeld1923}, but did not publish his results due to the criticism of Bohr and W. Pauli about the concept of non-integer quantum numbers. However, it was quite soon recognized that Heisenberg' success was accidental: similar attempts to describe excited states of the helium \cite{Born1923}, indeed, failed. It then became clear that ``some radical modification in the ordinary conceptions of the quantum theory or of the electron may be necessary'' \cite{VanVleck}.

The failure of the old quantum theory to describe successfully two-electron atoms, in fact, triggered (at least in part) the development of quantum mechanics in 1920s \cite{Mehra} and, once the basic formalism had been established by Heisenberg and E. Schr\"odinger, early variational calculations (the problem couldn't be solved analytically, as instead for the hydrogen atom) produced remarkably good results for the ground state of the helium atom, thus breaking the ground for the general acceptance of quantum mechanics.

In the present paper we review the known early results about this topic (Sect. \ref{s2}), and then discuss unknown results obtained almost simultaneously by the Italian physicist Ettore Majorana (Sect. \ref{s3}). As we will see, a large part of Majorana's results were deduced by making recourse to novel methods not yet appeared in the literature (both of that time and in present day studies), and while part of those numerical results were (and are) inaccurate when compared with the experimental data, the novel methods reveal quite useful in the frontier research related to atomic and nuclear physics. Finally, in Sect. \ref{s4} we give our conclusions and outlook.

\section{Solutions to the helium atom problem}
\label{s2}

\noindent Within the Schr\"odinger framework of quantum mechanics, atoms with two electrons are described by a wavefunction $\psi$ satisfying the eigenvalue equation (in electronic units):
\be \label{SE}
H \psi \equiv - \la \psi  + 2 \left( - \frac{Z}{r_1} - \frac{Z}{r_2} + \frac{1}{r_{12}} \right) \psi = W \psi \, ,
\ee
where $r_1, r_2$ are the distances of the first and second electron from the nucleus, $r_{12}$ their mutual separation and $W$ the energy of the system in Rydberg; $\la = \la_1 + \la_2$, with $\la_1$, $\la_2$ the Laplacian operators in the space of the first and second electron, respectively. This differential equation is not separable, so that, unlike the solutions for the hydrogen atom, its solutions for the eigenfunctions and energy eigenvalues cannot be expressed in closed analytic form, and thus approximation methods must be used. As it became clear soon, the key point where different strategies could be tested (thus proving or disproving the underlying theory) was the prediction about the ground state energy - mainly for the neutral helium atom - so that in the following we focus on this issue. Quite an exhaustive, though apparently outdated, review of the complete two-electron problem may be found in Ref. \cite{Bethe}. For future reference, it is useful to recall the experimental value of the ionization energy $W_I = -Z^2-W$ (here it is sufficient the accuracy as in late 1920s - early 1930s):
\be \label{IE}
W_I^{\rm exp} = 24.46 \, {\rm eV} \, .
\ee

\subsection{Perturbative calculations}  
\label{2.1}

\noindent The most simple method to solve Eq. (\ref{SE}) is to find a soluble problem to serve as an unperturbed solution to which the theory of perturbations may be applied. This was carried on by Sommerfeld's student A. Uns\"old in 1926 \cite{Unsold}, by factorizing the wavefunction in terms of spherical coordinates. The main idea and the basic calculations may be more easily understood by following a bit later derivation by Majorana, now reported in Ref. \cite{Quaderni}. In first approximation, the interaction among the two electrons may be neglected, so that the eigenfunction of the unperturbed hamiltonian,
\be \label{H0}
H_0 = - \la \psi  - 2 \left( \frac{Z}{r_1} + \frac{Z}{r_2} \right) \, ,
\ee
may be expressed as a product of two hydrogenic 1s states,
\be \label{psi0}
\psi = {\rm e}^{- Z r_1} {\rm e}^{- Z r_2} \, ,
\ee
with energy $W=-2 Z^2$. The inter-electron potential is, then, treated as a perturbation: by applying standard computational techniques, the first order result for the ground state energy (or ionization energy) is the following:
\be \label{first}
W = - 2 Z^2 + \frac{5}{4} Z \qquad \qquad \left[ W_I = Z^2 - \frac{5}{4} Z \right] .
\ee
For the neutral helium and ionized lithium, this translate in the predictions $W_I^{\rm He}=20.3$ eV and $W_I^{\rm Li} = 71.0$ eV, respectively \cite{Unsold}.\footnote{Or, more precisely, according to Majorana, $W_I^{\rm He}=20.31$ eV and $W_I^{\rm Li} = 71.08$ eV.} As noted by J.C. Slater few months after the appearance of the Uns\"old results \cite{Slater1927}, ``the solutions so far obtained, though qualitatively interesting, are quantitatively far from the truth'' when compared (for example, for the helium) with the experimental result (\ref{IE}). This urged the same Slater to search for an alternative strategy to the problem, by changing the starting point (that is, the unperturbed system) of the perturbation theory. The motivation is, apparently, mathematical in nature:
\begin{quote}
The method so far used in such cases has been to leave out, for the unperturbed problem, all those terms in the energy which prevent the separation of variables, regarding all these as perturbative terms, and solve the remaining problem by separation. Since these are, in general, the terms representing the interaction between electrons, it is obvious that the unperturbed problem will be a very poor approximation to the real case. We proceed in quite a different way, making no attempt to separate variables at all \cite{Slater1927}.
\end{quote}
The idea was to treat the variables of one of the two electrons as {\it parameters} in the Schr\"odinger equation of the other electron (that is, the equation for an electron in the field of the rest of the atom). From a more physical perspective, Slater imagined one electron in a fixed position and then obtained for the second electron a two-center problem, the eigenfunctions of which could be evaluated numerically as function of the distance between the two centers. For the first electron he deduced, thus, a numerical equation in which the eigenvalue of the remaining electron was introduced and which represented the motion of this electron in a central field. Then, both solutions were integrated into an approximate solution for the two-electron problem, and finally a perturbative calculation was performed. By using, initially, poorly accurate solutions to the unperturbed problem, Slater found a value $W_I^{\rm He}=24.35$ eV, apparently very close to the experimental determination, but in a subsequent note added in proof \cite{Slater1927} he corrected such a prediction, by obtaining a numerical value of $5 \div 6 \%$ greater than the experimental value (which was, however, a better result when compared with the Uns\"old prediction). Almost a year later, Slater published a paper \cite{SlaterA} where he analyzed the reasons of why his perturbative method got worse with increasing perturbative order, despite the outer electron in the helium atom was assumed to move not in a hydrogen-like
field, as for Eq. (\ref{psi0}), but in a certainly more accurate central field. The wavefunction replacing that in Eq. (\ref{psi0}) (for the helium problem) for this purpose was the following:
\be \label{psislater}
\psi = {\rm e}^{-2(r_1+r_2) + r_{12}} \, .
\ee
Slater found that his method indeed gave a good approximation to the energy levels and to the perturbed wavefunction, except at very small distances. More in general:
\begin{quote}
Almost any method of setting up approximate wavefunctions will give fairly good results in the non-penetrating part of the orbit; and it is this part that is of almost complete importance in many practical applications that demand knowledge of the wavefunction [...] The operator $H$, almost alone among the operators of physical interest, puts a sever strain on the most inaccurate part of the wavefunction, that connected with the penetrating part of the orbit, on account of the differentiation and the factors $1/r$ that it contains. For this reason, a wavefunction which is decidedly satisfactory for most physical applications may give a very inaccurate matrix of $H$. Yet it is just this matrix that is used in the perturbation method for solving the atomic structure problem. It seems necessary to conclude from this that the perturbation method is not a satisfactory one for actual calculations \cite{SlaterA}.
\end{quote}
A different method, then, appeared to be required for improving the predictions of quantum mechanics about two-electron atoms.\footnote{Note, however, that Slater applied his conclusions to any atom.}

\subsection{Ritz variational method}  
\label{2.2}

\noindent Such a novel method came into the play a bit earlier by means of G.W. Kellner \cite{Kellner}, a student of M. von Laue in Berlin, who introduced for the first time the variational Ritz method \cite{Ritz} in order to estimate the ground state energy of the helium atom. This is given by the minimum value of the Schr\"odinger variational integral
\be \label{min}
W[\psi] = \frac{\dis \int \psi^\ast H \psi \, {\rm d}\tau}{\dis \int \psi^\ast \psi \, {\rm d}\tau}
\ee
with respect to any wavefunction $\psi$, but, when considering only a given set of functions $\psi$, the minimum will correspond to an approximate value, which is better and better as the set is enlarged. When this set reduces to the unperturbed wavefunction in Eq. (\ref{psi0}), the variational method gives the same result for the ground state energy seen above, so that a larger set of functions is required for a better approximation.
This is achieved by choosing a trial wavefunction depending on some arbitrary parameters, the values of which are determined by requiring the energy functional in Eq. (\ref{min}) to be the lowest possible. This  minimum value then gives an upper limit for the ground state energy, whose approximation is much better as the trial wavefunction is close to the ``true'' eigenfunction of the problem, a suitable form for that being set up from given physical considerations. In principle, by including a sufficient number of arbitrary parameters,  one can approximate the eigenfunction and eigenvalue of the problem as closely as desired, but it is clear that an appropriate choice of the form of the wavefunction will make the procedure converge rapidly. 

Kellner introduced the novel idea of an {\it effective} nuclear charge experienced by one of the two electrons in the field of the nucleus {\it and} of the other electron: this electron somewhat shields the nucleus, thus reducing the effective charge. This quantity was introduced as an arbitrary parameter in the trial wavefunction replacing that in Eq. (\ref{psi0}). By following, again, the simpler (and somewhat more general) later reworking by Majorana \cite{Quaderni}, the wavefunction thus takes the form
\be  \label{psi1}
\psi = {\rm e}^{- k (r_1 + r_2)} 
\ee
and, minimizing the energy functional in Eq. (\ref{min}), the following value of the $k$ parameter is obtained:
\be \label{eff}
k = Z - \frac{5}{16} \, .
\ee
For the helium atom, then, the effective nuclear charge is less than two, $k_{He} = 27/16$, corresponding to a ionization energy $W_I^{\rm He} = 217/128$ (in Rydberg units) or $W_I^{\rm He} = 22.95$ eV which is in a better agreement with experiments when compared with the values obtained with perturbation methods. Majorana \cite{Quaderni} also found a general formula for the ground state energy (or ionization energy) by using the present method:
\be \label{EMvar}
W = - 2 \left( Z - \frac{5}{16} \right)^2 \qquad \qquad \left[ W_I = Z^2 - \frac{5}{4} Z  + \frac{25}{128} \right].
\ee
Instead, turning back to the helium atom, Kellner \cite{Kellner} also re-iterated the method, by using more involved wavefunctions, where the exponential factor in (\ref{psi1}) was multiplied by different combinations of Laguerre polynomials (in $r_1$ and $r_2$) and spherical harmonics (in the cosine of the angle between $r_1$ and $r_2$). After four approximation steps, the better result he obtained was $W_I^{\rm He} = 23.75$ eV, thus showing the Ritz variational method to be suitable for accurate quantum mechanical treatment of the ground state of helium, depending on appropriate choices for the wavefunctions.

\subsection{Hartree's self-consistent field method}  

\noindent A different idea in managing, in general, many-electron atoms was introduced by D.R. Hartree at the end of 1927 \cite{Hartree}. He started by assuming the wavefunction of the given atom to be written in the form of a product of single particle wavefunctions $y_i(r_i)$, one for each of the electrons, such as, for example, the one in (\ref{psi0}) for the helium atom. The potential energy acting on the first electron is taken to be the Coulomb field of the nucleus corrected by the potential of the charge distribution $|y_2(r_2)|^2$ of the second electron, averaged over all positions of the second electron:
\be \label{potHart}
V_1(r_1) = - \frac{Z}{r_1} + \int \frac{|y_2(r_2)|^2}{r_{12}} \, \drm \tau_2 \, .
\ee
This corrected {\it central} potential was, then, substituted into the Schr\"odinger equation for the first electron and solved for the wavefunction $y_1(r_1)$. A similar procedure is carried out for the second electron and, from the solutions $y'_i(r_i)$ for all electrons, Hartree calculated a distribution of charge to be inserted again in Eq. (\ref{potHart}) to obtain the field of the nucleus and this charge distribution. In general, this ``final'' field is not the same as the ``initial'' field, since the initial, trial wavefunctions $y_i(r_i)$ do not agree with the final ones $y'_i(r_i)$. The whole procedure is, then, repeated by using the final field of the first approximation as the initial field of the second one, and this is repeated over and over again until the initial and final wavefunctions agree to the desired accuracy. The final field is thus the same as the field produced by the charge distribution of the electrons, so that the field in which the electrons are assumed to move when calculating atomic quantities of interest (such as the ground state energy) is  ``self-consistent''.

Here, Hartree's basic assumption was that many-electron atoms could be described by wavefunctions given as the simple product of one-electron wavefunctions of individual electrons, rather than as one complete wavefunction (as, for example, that in Eq. (\ref{psislater})). This was criticized by J.A. Gaunt \cite{Gaunt} and others but, although Hartree's method  ``seemed to many persons to stand rather apart from the main current of quantum theory'' since it  ``seemed to contain arbitrary and empirical elements'' \cite{SlaterNote}, Gaunt admitted that  ``Hartree's wavefunctions have been shown to be good approximations'' \cite{Gaunt}. This was clarified a bit later by Slater \cite{SlaterB}, who evaluated the error bars of Hartree's results, by making a (general, not limited to helium) perturbative calculation of the energy levels by using the wavefunction obtained with the Hartree method.

Hartree applied his method to Rb, Rb$^+$, Na$^+$, Cl$^-$ and, first of all, to helium, obtaining a value for the ionization potential, $W_I^{\rm He} = 24.85$ eV, very close to the experimental determination (\ref{IE}).

In passing, we also mention a different, independent attempt made by Slater \cite{SlaterC} to find an accurate expression for the wavefunction of the ground state of \mbox{(ortho-)} helium, which was a further refinement of previously considered Eq. (\ref{psislater}). The general idea was to use theoretical determined functions for the limiting cases of large and small $r$-distances and then interpolate between them. In particular, when one electron is a considerable distance from the nucleus (large $r_2$), while the other close up (small $r_1$), the wavefunction is assumed to be the product of an ionic function of the inner electron and a hydrogen-like function of the outer one:
\be \label{psislaterC1}
\psi = {\rm e}^{-2 r_1} \frac{\erm^{- a r_2}}{r_2^b} \left( 1 + \frac{c}{r_2} + \dots \right)
\ee
(the opposite case, with small $r_2$ but large $r_1$, is obtained by interchanging $r_1$ and $r_2$ in this expression). Instead, when both electrons are close up, the wavefunction is assumed to be of the same form as in (\ref{psislater}), but with a further correction term:
\be \label{psislaterC2}
\psi = {\rm e}^{-2(r_1+r_2) + (1/2) r_{12} + d (r_1^2 + r_2^2)} \, .
\ee
The novel term weigthed by $d$ was added by Slater in order to have a smooth interpolation between the two expressions in Eq.s (\ref{psislaterC1}) and (\ref{psislaterC2}), for intermediate values of $r_1, r_2$. Note that such wavefunctions were {\it not} intended for evaluating the energy eigenvalues, since the coefficients $a,b,c,d$ were just deduced from the experimental spectroscopic data. Instead, they were introduced in order to compute several other atomic properties of helium, such as diamagnetic susceptibility, repulsive forces between two helium atoms, and so on.
According to Slater \cite{SlaterC}:
\begin{quote}
The calculations were made before the writer saw Hartree's paper, in which he obtains a charge density distribution for helium in quite a different way. The wavefunction found in the present paper is more complicated than Hartree's in the matter of the way in which it takes the interaction energy between electrons into account. But the charge density can be computed equally well from either method, and this permits a comparison of the present results with Hartree's. The discrepancies between the two are nowhere greater than one or two percent. This is highly satisfactory, both in that it verifies the present method and Hartree's, and also that it justifies us in believing this density distribution to be correct within a narrow limit of error. The other numerical results are also gratifying.
\end{quote}
In a sense, the trial stage of the novel theory of quantum mechanics was completed, and the subsequent need was just to have a  ``standard'' (and mathematically reliable for numerical purposes) procedure for evaluating atomic quantities to the desired accuracy. For many-electron atoms, this was just found in the Hartree method, 
as modified by V. Fock \cite{Fock} to take exchange effects into account (fully symmetrized wavefunctions in the form of products). Instead, for helium, the story was somewhat different.

\subsection{Hylleraas variables}  
\label{2.4}

\noindent Early in 1928, important novelties came into play in the variational approach to the ground state of the helium atom from E.A. Hylleraas, a Norwegian student of M. Born in Gottingen \cite{HylRemi}. The variational method already exploited by Kellner was adopted because, according to Hylleraas, a ``useful feature'' of it was that ``the eigenvalues obtained in successive orders decrease monotonically, and that therefore an exact eigenvalue can be given on the approach from one side'' \cite{Hyl1}. 
In the first of a series of papers devoted to the subject, he soon noticed \cite{Hyl1} that the wavefunction of the two-electron problem, which, in general, would depend on the six coordinates ${\mathbf r_1}, {\mathbf r_2}$, actually depended only on a subset of three, namely $r_1, r_2, \theta$ ($\theta$ being the angle between the two vectors ${\mathbf r_1}, {\mathbf r_2}$), defining the shape of the electron-nucleus triangle. Its absolute orientation in space was, indeed, out of interest for the problem at hand. This introduced an enormous simplification into the variational problem: indeed, Hylleraas himself was able to perform computations till to order 11 (instead of order 4 calculations performed by Kellner), thus obtaining an excellent approximation for the ionization energy of helium, $W_I^{\rm He} = 24.35$ eV.

The predictions by Kellner \cite{Kellner}, Slater \cite{Slater1927} and Hylleraas \cite{Hyl1} agree within the  error bars for direct measurements (by electron collision), but they resulted to be not precise enough in comparison to much more precise spectroscopic measurements by T. Lyman \cite{Lyman}. Hylleraas was well aware of this fact:
\begin{quote}
The end result of my calculations [...] was greatly admired and thought of as almost a proof of the validity of wave mechanics, also, in the strict numerical sense. The truth about it, however, was, in fact, that its deviation from the experimental values by an amount of one-tenth of an electron volt was on the spectroscopic scale quite a substantial quantity and might as well have been taken to be a disproof \cite{HylRemi}.
\end{quote}
For this reason, he continued to search for possible improvements in the subsequent years.

In a second paper \cite{Hyl2} on the topic, instead of adopting the $r_1, r_2, \theta$ coordinates, he chose only ``metric quantities appearing immediately in the expression for the potential energy", that is  $r_1, r_2, r_{12}$. In particular, he introduced the notations (later known as ``Hylleraas variables'')
\be \label{hylvars}
s= r_1 + r_2, \qquad \qquad t = r_2 - r_1, \qquad \qquad u = r_{12},
\ee
so that the wavefunction had to be written just in terms of these coordinates. The variational method was then applied to the function $\psi = \varphi (ks, kt, ku)$ - that is, with $k$ as a variational parameter - with
\be \label{phihyl}
\varphi (s,t,u) = \erm^{- s/2} P(s,t,u), 
\ee
$P(s,t,u)$ being a polynomial in $s,t,u$ variables. Note that already Kellner had tried to force a faster convergence by multiplying the metric arguments in his wavefunctions by a free constant, but here the novelty was in the choice of the variables in (\ref{hylvars}). Very accurate results were obtained, depending - obviously - mainly on the number of terms included in $P(s,t,u)$: increasing the number of accurately chosen polynomial terms resulted in better approximations. The method worked very fine for parahelium (and a little less fine for orthohelium) and, after several trials with different monomial terms $s^\ell t^m u^n$ (different values of $\ell,m,n$), Hylleraas was able to obtain the value of  $W_I^{\rm He} = 24.460$ eV with order 6 calculations \cite{Hyl2}. 

Soon after the appearance of \cite{Hyl2}, H.A. Bethe compared this result with that obtained with the Hartree method, showing that the overall charge distribution of the two electrons given by the Hartree self-consistent field method (with correlation effects not included) agreed very well with the predictions of the variational method \cite{Bethe1929}.

Later in 1929, Hylleraas extended \cite{Hyl3<} his previous calculations \cite{Hyl1}, \cite{Hyl2} to helium-like ions Li$^+$ and Be$^{++}$:
\begin{quote}
A most exciting cooperation now started between [me and the famous spectroscopist Bengt Edl\'en]. No sooner had I found the energy of the lithium ion lying in between his limits of error than the doubly ionized beryllium ion was on the way, and so it continued with boron 3 plus and carbon 4 plus with me always behind. Suspecting there were no limits to Edl\'en's power of producing ions I decided to overtake him with a good safety margin by introducing even an infinite nuclear charge. This was the origin of the energy formula $E = - 2 Z^2 + \frac{5}{4} Z - \epsilon_2 + \epsilon_3/Z - \epsilon_4 / Z^2 + \ldots$ as counted in Rydberg units \cite{HylRemi}.
\end{quote}
Hylleraas considered the term $1/Z r_{12}$ in the potential energy as a perturbation function, and the factor $1/Z$ as a variational parameter. The energy eigenvalue $E$ and eigenfunction $\psi$ were then assumed to be of the forms:
\bea
& & E = E_0 Z^2 + E_1 Z + E_2 + E_3 \, \frac{1}{Z} + \ldots , \\
& & \psi = \psi_0 + \psi_1 \, \frac{1}{Z},
\eea
and the following result for the ionization potential as function of $Z$ was obtained:
\be \label{WJHyl3}
W_I = Z^2 - \frac{5}{4} Z + 0.31455 - \frac{0.0147}{Z} .
\ee
It is interesting to compare the fully numerical evaluation leading to the last two terms in Eq. (\ref{WJHyl3}) with the corresponding analytic (and less accurate) evaluation in (\ref{EMvar}), accounting for a constant $25/128 \simeq 0.1953$ term. For the helium ionization energy there is no particular improvement carried on by this method with respect to that in \cite{Hyl2}, but here Hylleraas pushed the calculations till to order 10, with a prediction of  $W_I^{\rm He} = 24.469$ eV, to be compared with the Lyman experimental result of $24.467$ eV \cite{Lyman}.

Given the success of the numerical calculations performed by Hylleraas, a more mathematical insight into the Ritz variational method, as modified by Kellner and, later, by Hylleraas with the introduction of screening constants, was in order. First of all, in 1930, G. Breit showed that, due to the spherical symmetry of the field, the reduction of the general 6-dimensional problem to a reduced 3-dimensional one could always be performed for $S$-states. Then, few months later, C. Eckart proved that the quantity $E=\int \psi^\ast H \psi \, \drm \tau$, where $H$ is the negative energy operator, was a lower limit to the term-value of the lowest level of a given spectral series, and that the best approximation to that term corresponded to the largest value of the integral evaluated over a set of various $\psi$ function. He also showed, however, that such approximation was not so good at large distances from the nucleus. 

In the meanwhile, Hylleraas continued to improve his method, and to what anticipated in \cite{Hyl3<} he gave its final form in \cite{Hyl4}. Now, the trial wavefunction was written as
\be \label{psiHyl4}
\psi = \erm^{- \sigma/2} P(\sigma, \tau, \nu),
\ee
with screened arguments:
\be
\sigma = 2 Z s, \qquad \tau = 2 Z t, \qquad \nu = 2 Z u .
\ee
Eq. (\ref{WJHyl3}) was, then, replaced by the more accurate one:
\be \label{WJHyl4}
W_I = Z^2 - \frac{5}{4} Z + 0.31488 - \frac{0.01752}{Z} + \frac{0.00548}{Z^2} ,
\ee
with very small numerical coefficients for negative power of $Z$. In particular, calculations of the helium ionization potential were made more precise by three orders with respect to the previous estimate, thus obtaining the value $W_I^{\rm He} = 24.470$ eV.

\

The excellent agreement with experiments was very important in putting the quantum mechanical approach to atomic spectra, based on many-body Schr\"odinger equation, on a firm, quantitative ground. As it is apparent from the latest results discussed above, the matter rapidly converted from a theoretical problem to a numerical one and, in the subsequent years, the increasingly precise studies on two- or multi-electron atomic spectra developed into a small industry. The inspection of the subsequent editions of the Bethe and Salpeter book \cite{Bethe}, from 1933 to 2008, may give a good idea of the evolution of the subject.

\section{Majorana contributions: theory and numbers}
\label{s3}

\noindent Majorana was involved in helium in at least two occasions, namely for his second \cite{EM2} and third \cite{EM3} paper published in 1931, but both papers discuss {\it subsequent} applications of the helium atom, not directly related to what considered here. Nevertheless, they also contains key summary remarks worth to be reported.

Following the seminal work by W. Heitler and F. London \cite{Heitler} on the formation of the H$_2$ molecule, Majorana \cite{EM2} considered as well the problem of the chemical bond, applied to the study of the more intriguing case of the molecular ion He$^+_2$ \cite{Pucci}. Such a paper gave the first quantitative description of this molecular ion within the framework of the Heitler and London method (and whose results agreed with the available experimental data), but, for what concerns us here, it is somewhat interesting for the summary made by the author about the wavefunctions for the helium atom, to be employed in calculations of observable quantities.
\begin{quote}
The eigenfunction of the neutral atom of helium in its ground state has been calculated numerically with great accuracy but does not have a simple analytical expression \cite{SlaterC, Hyl2}. Therefore we need to use rather simplified unperturbed eigenfunctions. For example, we could assume, as commonly done, for the helium ground state the product of two hydrogen-like eigenfunctions with an effective $Z$ equal to $1.6 \div 1.7$, depending on the criterium used for the evaluation; if we want to optimize (minimize) the average energy, we must then set $Z = 2-5/16 = 1.6875$; if instead we want that the diamagnetic constant be in agreement with the experimental value and at the same time with the value provided by accurate theoretical calculations \cite{SlaterC}, we must set $Z = 1.60$ \cite{EM2}.
\end{quote}
It was customary for Majorana to relate his theoretical calculations to experimental observations \cite{EMbio}, and then to adopt appropriate approximations for the former in order to fit the accuracy of the latter, so that such a summary well depict the situation around 1930. However, this also explains his reiterated use of hydrogenoid wavefunctions and, in particular, his constant search for their possible generalizations giving easy but physically meaningful expressions (see below).

A second paper, published in the same year \cite{EM3} (but anticipating of few years what later discovered by other authors \cite{EM3others}), dealt with the calculation of certain doubly excited levels of helium within a variational perturbative approach. Here Majorana gave a further generalization of his (unpublished) formula reported in Eq. (\ref{EMvar}), also somewhat reminiscent of some empirical relations found earlier and discussed below (see Sect. \ref{3.2}). The reasoning was as follows. In the zeroth approximation (with respect to the mutual interaction between the two electrons), the ground state energy (in Rydberg) is given by $W= - 2Z^2$, while in the first approximation it will be $W= - 2Z^2 + 4 a Z$ since the interaction increases as $Z$. Now, 
\begin{quote}
the second approximation can be evaluated with the method of the variation of the unit of length \cite{Fock63}\footnote{Majorana refers simply to the Hylleraas method (see Sect. \ref{2.4}); the paper by Fock quoted here deals with just a general subject, i.e. the virial theorem in the framework of quantum mechanics, where the Hylleraas method is mentioned as an example.} that is here equivalent to assuming hydrogen-like eigenfunctions with an effective $Z^\ast$ \cite{EM3}. 
\end{quote}
The ``more correct expression'' he got with this method was the following:
\be \label{EM729}
W= - 2Z^2 + 4 a Z - 2 a^2 ,
\ee
the screened charge being $Z^\ast = Z-a$, that is, just as in Eq. (\ref{EMvar}) but with arbitrary $a$.\footnote{Note that in Ref. \cite{EM3}, Majorana applied this same formula to the ground state as well as to other levels (with different values of $a$).} An explicit comparison of what Majorana obtained with the method mentioned with the available results was performed too, thus showing the goodness of his estimate:
\begin{quote}
With this method we find that the value of the ground state is $- W/{\rm Rh} = 729/128 = 5.695$, whereas the empirical value and the theoretical one obtained by Hylleraas is $- W/{\rm Rh}= 5.807$, the difference being less than $2\%$ \cite{EM3}.
\end{quote}

The work peformed by Majorana behind the results he presented in his papers (and not published), however, is far from obvious. Indeed, in his unpublished personal study \cite{Volumetti} and research \cite{Quaderni} notes we find several interesting results and methods directly related to the two-electron problem, dating back to 1928-9, some of which are apparently preliminary studies for his published papers (especially Ref. \cite{EM3}). In the following we give an account of what obtained by Majorana in those years, by sketching some key points. 

Although we do not know precisely when and how Majorana approached the two-electron problem (with particular reference to helium), it is a matter of fact that in his personal notes we find both purely theoretical contributions and numerical calculations, including empirical relations as well. The unpublished material we present below does not necessarily follow closely the time evolution of Majorana's reasoning.

\subsection{A general variant of the variational method}


An interesting modification of the variational method was the following \cite{Quad130-1}. Majorana considered an arbitrary function $\varphi$ in terms of which the wavefunction of the ground state was written as
\be \label{EMpsitep3}
\psi = a_0 \varphi + a_1 H \varphi ,
\ee
$H$ being the Hamiltonian operator. The two variational parameters $a_0,a_1$ have to be determined  (for a given $\varphi$) by minimizing the energy functional in (\ref{min}), which assumes the form
\be
W_{} = \frac{a_0^2 A_1 + 2 a_0 a_1 A_2 + a_1^2 A_3}{a_0^2 + 2 a_0 a_1 A_1 + a_1^2 A_2}, 
\ee
with
\be \label{AN}
A_n = \frac{\dis \int \varphi^\ast H^n \varphi \, {\rm d}\tau}{\dis \int \varphi ^\ast \varphi \, {\rm d}\tau} 
\ee
($n=1,2,3$). By introducing the function $f(a_0,a_1) = a_0^2 A_1 + 2 a_0 a_1 A_2 + a_1^2 A_3 = W_{} a_0^2 + 2 W_{} a_0 a_1 A_1 + W_{} a_1^2 A_2$, from the minimizing conditions  $\prt f / \prt a_0 =0$, $\prt f / \prt a_1 =0$ the set of equations for deducing $a_0, a_1$ is obtained:
\be
\ba{l} \dis
a_0 (A_1 -W_{}) + a_1 (A_2 - W_{} A_1) = 0, \\ \dis 
a_0 (A_2 - W_{} A_1) + a_1 (A_3 - W_{} A_2) = 0 .
\ea 
\ee
Now, this is a homogeneous set of equation which admits non trivial solutions for $a_0, a_1$ only by requiring the matrix of coefficients to have a vanishing determinant:
\be 
\left| \begin{array}{ll}
A_1 - W & A_2 - A_1 W \\
A_2 - A_1 W & A_3 - A_2 W \end{array} \right|  = 0. 
\ee
This last condition allows to determine the energy $W$ as the smallest root of the corresponding equation, that is
\be
W = \frac{A_3 - A_1 A_2 + \sqrt{(A_3 - A_1 A_2)^2 - 4 (A_1 A_3 - A_2^2)(A_2 - A_1^2)}}{2(A_2 - A_1^2)} .
\ee
In the {\it Quaderni} Majorana did not apply such a method to come up with numerical predictions about the ground state energy of helium, but it is nevertheless interesting to compare such predictions with already known results. For example, by choosing the form of $\varphi$ as that in (\ref{psi1}), we have obtained 
$W_I^{\rm He} =21.62$ eV (for $k=1.6711$), which is certainly not a good result when compared already with what obtained by Majorana himself with the standard variational method for the same trial wavefunction (see Sect. \ref{2.2}), i.e. $W_I^{\rm He} =22.95$ eV. 

This is probably the reason why Majorana further generalized his method, by writing the wavefunction:
\be \label{EMgenvar1}
\psi = a_0 \varphi + a_1 H \varphi + a_2 H^2 \varphi + \ldots + a_n H^n \varphi
\ee
for a given $\varphi$ and a general index $n$, for which the energy to be minimized is (with the same notations as above):
\be
W = \frac{{\dis \sum_{i,k=0}^n a_i
a_k \ A_{i+k+1}}}{{\dis \sum_{i,k=0}^n a_i a_k \ A_{i+k}}}.
\ee
``$W$ will be the smallest root of the following equation:"
\be \label{EMgenvar2}
\left| \begin{array}{lllll}
A_1 - W & A_2 - A_1 W             
& \ldots &  &  A_n -A_{n-1} W \\
A_2 - A_1W & A_3 - A_2 W          
& \ldots &  &  A_{n+1} -A_n W \\
A_3 - A_2 W & A_4 - A_3 W         
& \ldots &  &  A_{n+2} -A_{n+1} W \\
\ldots   &  &  &  &    \\
A_n - A_{n-1}W & A_{n+1} - A_n W  
& \ldots &  &   A_{2n-1} -A_{2n-2} W \end{array} \right| = 0 .
\ee
Obviously, by increasing $n$, the calculations give better and better results but, as well, become more and more difficult. However, the point on which Majorana focused was not that but, rather, the emergence of divergencies:
\begin{quote}
Often, this procedure does not converge, because, starting from a given value of $n$, quantity $H^n \varphi$ exhibits too many singularities, which forces us to consider only combinations of the form
\be  \label{EMnstop}
\psi = a_0 \varphi + a_1 H \varphi + \ldots + a_{n-1} H^{n-1} \varphi. 
\ee 
The inclusion of additional terms is not useful, since the corresponding $a$ coefficients would necessarily vanish.
\end{quote}
He referred to the calculations of integrals for $r_1=0$, $r_2=0$ or $r_{12}=0$, whose degree of divergency increases with incrasing $n$ in $H^n$, given the form of the Hamiltonian operator. It is quite interesting the connection made by Majorana between the form (for a given $\varphi$) of $\psi$ in (\ref{EMnstop}), dictated by choosing a suitable $n$, and the divergency properties of the system, as described by its Hamiltonian.\footnote{Recall that such contribution dates around 1928-9.}

\subsection{Empirical relations}
\label{3.2}

Seemingly, the first numerical studies on the two-electron problem performed by Majorana were aimed to find empirical relations for the ground state energy \cite{Vol267}. He was evidently aware of the difficulties in obtaining theoretical predictions,\footnote{Apart from the references in \cite{EM2} and \cite{EM3}, we do not know precisely what papers were read and studied by Majorana, but from the arguments and, especially, from the symbols he used in his notes \cite{Volumetti, Quaderni}, we can safely deduce that he was well aware of a large part of the literature quoted above.} so that the first numerical efforts were likely devoted to find suitable empirical expressions for $W$. According to Majorana himself, several of these relations lead to unsatisfactory results, but it is nevertheless intriguing the reasoning behind them.\footnote{In what follows, we use consecutive labels for the equations in the quotations from the original Majorana's manuscript; we also conform, when necessary, the original notation to the present one.}
\begin{quote}
Let us consider a two-electron atom with charge $Z$ in
its ground state. We denote by $a \, = \, < \! 1/r_1 \! > \, = \,
< \! 1/r_2 \! >$ the mean value of the inverse of the distance of
each electron from the nucleus, and with $b = <1/r_{12}>$ the mean
value of the inverse of the distance between the two electrons.
Expressing the distances in electronic units and the energy in
Ry, we have
\begin{equation}\label{v3-15-1}
  W \; = \; - 2 \, a \, Z \, + \, b  ,
\end{equation}
since the energy is equal to half the mean value of the potential energy.
If we now consider an atom with atomic number $Z + dZ$, perturbation theory
gives
\begin{equation}\label{v3-15-2}
  \drm W \; = \; - \, 4 \, a \, \drm Z ,
\end{equation}
and thus we have two equations in the three unknown $Z$ functions $E,a,b$ \cite{Vol267}.
\end{quote}
Interestingly enough, Majorana's strategy for dealing with generic two-electron atoms with atomic number 
$Z$ was to consider this quantity as a {\it continuous} variable, a small change of which induces a change in the ground state energy, the latter one being evaluable with the aid of the perturbation theory. Now, the problem was to have two equations with three unknowns, so that the necessary additional relation was introduced as an empirical formula:
\begin{quote}
We now add another empirical relation between $a$ and $b$, which is
presumably a good approximation:
\begin{equation}\label{v3-15-3}
  b \; = \; ( 2 Z \, - \, 2 a ) \, ( 2 a \, - \, Z )  .
\end{equation}
This relation can be deduced from the following considerations. For
sufficiently high values of $Z$, perturbation theory gives
\begin{equation}
  W \; = \; - 2 \, Z^2 \, + \, \frac{5}{4} \, Z \, + \, \dots ;
\end{equation}
but, on the other hand,
\begin{equation}
  b \; = \; \frac{5}{8} \, Z \, + \, \dots ,
\end{equation}
so that, from Eq. (\ref{v3-15-1}),
\begin{equation}
  a \; = \; Z \, - \, \frac{5}{16} \, + \, \dots  ,
\end{equation}
which, in first approximation, satisfies Eq. (\ref{v3-15-3}). For very
small values of $Z$ we can consider that the first electron is next to the
nucleus, while the other one is practically at an infinite distance; then
we have $a \simeq Z/2$, $b \simeq 0$, and Eq. (\ref{v3-15-3}) is again
satisfied. We finally assume that it is also a good approximation for
intermediate values of $Z$.
\end{quote}
The next step was, then, simply to substitute Eq. (\ref{v3-15-3}) into Eqs. (\ref{v3-15-1}) and (\ref{v3-15-2}), thus obtaining $\drm Z = \drm a$, and ``since we know the value of $a$ for infinite $Z$, we deduce $a = Z - 5/16$.'' The ground state energy is, then, accordingly,
\be \label{ve1}
W = - 2 \, Z^2 \, + \, \frac{5}{4} \, Z \, - \, \frac{25}{64}  ,
\ee
to be compared with the perturbative result in Eq. (\ref{first}) and subsequent improvements. Majorana noticed that such formula could be used only for $Z \geq 5/8$, while for $Z=5/8$ the quantity $b$ in (\ref{v3-15-3}) vanishes. Moreover, ``the procedure used here is not very satisfactory, since for
very small $Z$ the quantity $b$ would vanish faster than a first-order term and it would become negative.''
A comparison with data for the helium atom revealed that the predicted value for $W$ was in excess of 1.13 eV with respect to the experimental value (25.59 eV instead of 24.46 eV), but here the intriguing thing is that Majorana applied the above formula also to hydrogen: ``For the hydrogen atom ($Z=1$) we find instead  $W = -1.141$, from which the ionization potential would be $1.91$ (electron affinity)."\footnote{Note that the predicted electron affinity of the hydrogen atom, i.e. the difference between the ground state energies of the neutral atom and the once-ionized atom, is more than twice the actual value.}

Given the poor agreement with the experimental values, Majorana also explored other relations for $b$ as functions of $Z$, such as, for instance, 
\be \label{emp1}
 b =  \frac{5}{8} \left( \sqrt{k^2 + Z^2} \, - \,  k \right) ,
\ee
or 
\be \label{emp3}
  b  =  \frac{5}{8} \left( \sqrt[3]{k^3 + Z^3} \, - \, k \right)  ,
\ee
or even
\be \label{emp2}
  b  = \frac{5}{8} \, Z \, {\rm e}^{-k/Z} , 
\ee
where $k$ is a parameter to be determined under suitable conditions. For example, from the last relation, he obtained the following expression for $a$:
\be
a = Z \left( 1 \, + \, \int_0^\infty \, \frac{5 Z + 5 k}{16
\, Z^3} \, {\rm e}^{-k/Z} \, \drm Z \right)  ,
\ee
where it is evident the contribution from {\it any} $Z$ ranging from $0$ to $\infty$.
\begin{quote}
Since for small $Z$ we must have $a \simeq Z/2$, we can choose $k$ such
that
\begin{equation} \label{v3-15-19}
\int_0^\infty \, \frac{5 Z + 5 k}{16 \, Z^3} \, {\rm e}^{-k/Z} \,
\drm Z \; = \; \frac{1}{2} ,
\end{equation}
i.e., $k = 5/4$. However, in this way we obtain a bad
approximation. Indeed, we would have [...] 
\be
E = - \, Z^2 \, - \, Z^2 \, {\rm e}^{-1.25/Z} , 
\ee 
and for helium ($Z=2$) we would get $E=-5.14$, which is a value far from the experimental one.
\end{quote}
Majorana was well aware that the introduction of such empirical relations was completely arbitrary (``there is no {\it a priori } reason to prefer one or the other''), so that he soon abandoned these calculations. Here, however, it remains the basic, novel assumption that $Z$ could be treated - in an ``effective'' way - as a continuous variable, and that contributions to the ground state energy of a given atom (with given atomic number) come also from any other $Z$.

\subsection{Helium wavefunctions and broad range estimates}
\label{3.3}

The conclusions achie\-ved by Slater in 1928 \cite{SlaterA} about the form of the unperturbed wavefunction and the role played by the operator $H$ in the perturbative method (see Sect. \ref{2.1}), were analyzed repeatedly by Majorana \cite{Quad133}, who considered as well several different forms of the helium wavefunction as generalization of (\ref{psi0}), along the same line considered by Slater (who introduced, in particular, the simple form in (\ref{psislater})). However, from the given wavefunctions, Majorana was able to deduce also (broad) ranges within which the value of the helium ground state energy would have lain, thus quantifying in a definite manner what generally deduced by Slater. 

Majorana realized very early what later employed extensively by Hylleraas (but already latent in Slater's calculations), namely that the helium wavefunction depends only on three metric quantities $r_1, r_2, r_{12}$ (see also Sect. \ref{3.4}). Probably inspired by Slater's Eq. (\ref{psislaterC2}), Majorana firstly introduced the following general form:
\be \label{EMw}
\ba{c}
\dps \psi = \erm^{-p} , \\
\dps p = \frac{\dps 2 r_1 + 2r_2 - \frac{1}{2} r_{12} + a (r_1^2 + r_2^2) + br_1 r_2 +
c r_{12}^2 + d (r_1 + r_2)r_{12}}{\dps 1 + e (r_1 + r_2) + f r_{12}} , 
\ea
\ee
where $a,b,\dots,f$ are numerical coefficients. The energy eigenvalue equation (\ref{SE}) was then written (in suitable units) as 
\be \label{EMlambda}
L \psi = \lambda \psi
\ee
with
\bea   
L & = &  \frac{4}{r_1} + \frac{4}{r_2} -
\frac{2}{r_{12}}+ \la \nonumber \\ &=& \frac{4}{r_1} + \frac{4}{r_2} -
\frac{2}{r_{12}} + \frac{\partial^2}{\partial r_1^2} +
\frac{\partial^2}{\partial r_2^2} + 2 \frac{\partial^2}{\partial
r_{12}^2} + \frac{2}{r_1} \frac{\partial}{\partial r_1} +
\frac{2}{r_2} \frac{\partial}{\partial r_2} + \frac{4}{r_{12}}
\frac{\partial}{\partial r_{12}} \nonumber \\
& & + 2 \cos \alpha_1 \cdot \frac{\partial^2}{\partial r_1
\partial r_{12}}
+ 2 \cos \alpha_2 \cdot \frac{\partial^2}{\partial r_2
\partial r_{12}}  , 
\eea
where $\alpha_{1}$[$\alpha_{2}$] is the angle between ${\mathbf r_1}$[${\mathbf r_2}$] and ${\mathbf r_{12}}$. However, he soon realized that calculations would become quite difficult with such a general form, so that he shifted his attention to the simpler form
\be \label{EMw1}
\psi_0 = \erm^{-2r_1 - 2 r_2 + \frac{1}{2} r_{12}} .
\ee
Though, it should be noted that such a choice was not accidental (or dictated only by computational simplicity): indeed, he deduced the relations to be hold among the $a-f$ coefficients in order to have from Eq. (\ref{EMw}) the same result (see below) as coming from Eq. (\ref{EMw1}).\footnote{For technical details, see \cite{Quad133}.} By substituting (\ref{EMw1}) into the eigenvalue equation (\ref{EMlambda}), Majorana obtained simply:
\be
\lambda = \frac{17}{2} - 2 \cos \alpha_1 - 2 \cos \alpha_2 .
\ee
From this expression, by allowing the cosines to take any value in their domain, he deduced the range inside which the (negative) energy eigenvalue has to lie, namely $4.5 \leq -W/{\rm Rh} \leq 8.5$. 

When compared with the value of 5.807, this is indeed a very broad range, but, as said above, it should give a quantitative idea of the results obtainable within the perturbative method, as induced by the particular form of the Hamiltonian operator on different unperturbed wavefunctions. Therefore, Majorana passed to consider an 
``approximated" wavefunction, that is
\be \label{EMw2}
\psi = \left( 1 + \frac{1}{2}r_{12} \right) \erm^{-2 r_1 -
2r_2}.
\ee
Note that such expression approximates that in (\ref{EMw1}) (and, then, that in (\ref{EMw})) for small $r_{12}$, i.e. when the two electrons are close each other: it can be then regarded as a correction to the simple one-electron case. By repeating the same calculations as above, Majorana obtained
\be 
\lambda = 8 - \frac{1}{1 + \frac{1}{2} r_{12}} - \frac{2}{1 + \frac{1}{2}
r_{12}} \left( \cos \alpha_1 + \cos \alpha_2 \right) ,
\ee
and thus $3 \leq -W/{\rm Rh} \leq 8$. The comparison of this range with that obtained previously already quantifies to a certain extent the result obtained by Slater \cite{SlaterA} and addressed above, but Majorana did not limit himself to just this two forms of the possible wavefunctions. Without entering into computational details \cite{Quad133}, we only mention a different approximation of (\ref{EMw}), namely
\be
\psi = 1  - 2 r_1 - 2 r_2 + \frac{1}{2} r_{12} + a(r_1^2 + r_2^2) + b r_{12}^2 + c r_1r_2 +
d (r_1 + r_2) r_{12} + \ldots
\ee
for small $r_1, r_2, r_{12},$\footnote{In particular, Majorana deduced the numerical values for the coefficients by requiring that $\psi$ and its first derivative have a node at the same position when the two-electron system collapses into a one-electron one, i.e. $r_1=0$ (or $r_2=0$) and $r_{12}=0$.} and a generalization of (\ref{EMw2}),
\be
\psi = \left( 1 + \frac{1}{2} \ r_{12} \right) \left[ \frac{\erm^{-2 r_1- (2 - 2 \alpha) r_2}}{1 + 2
\alpha r_2} + \frac{\erm^{-(2-2 \alpha) r_1 - 2 r_2}}{1 + 2 \alpha r_1} \right]
\ee
(obviously symmetric under the exchange $1 \leftrightarrow 2$), with the parameter $\alpha$ parameterizing an effective nuclear charge {\it different} for the two electrons, probably due to the different position of the two electrons considered by Majorana with respect to the nucleus.

\subsection{A simpler alternative to Hylleraas' method}
\label{3.4}

As we have in Sect. \ref{2.4}, a sensible improvement in the predictions obtained with the variational method came with the recognition by Hylleraas that the wavefuntion of the two-electron problem depends only on three coordinates (instead of six), namely the magnitude of the two vectors ${\mathbf r_1},{\mathbf r_2}$ and the angle $\theta$ between them or, rather, $r_1, r_2, r_{12}$. In addition to the perturbative calculations just discussed, and likely before the appearance of the Hylleraas papers \cite{Hyl1, Hyl2} (or, in any case, independently of them), Majorana implemented such a result in one of his (unpublished) variational studies about the ground state energy of two-electron atoms \cite{quadtep4}, aimed at obtaining an improvement of his general result in Eq. (\ref{EMvar}). Instead of the wavefunction in (\ref{psi1}), he considered the function
\be \label{psitep4}
\psi = \erm^{- k r_1} \erm^{-k r_2} \erm^{\ell r_{12}}
\ee
with arbitrary $k$ and $\ell$ to be determined: ``we will certainly obtain a better approximation.'' The energy in (\ref{min}) was, then, evaluated by noting that:
\be
H\psi = - 2k^2\psi -2 \ell^2\psi + 2 k \ell\psi
a_{1} + 2 k\ell\psi a_{2} - 2 \frac{Z-k}{r_1}\psi - 2 \frac{Z-k}{r_2}
\psi + \frac{2 - 4\ell}{r_{12}}\psi ,
\ee
where $a_{1}$ [$a_{2}$] is the cosine of the angle between ${\mathbf r_1}$ [${\mathbf r_2}$] and ${\mathbf r_{12}}$. However, as subsequently done also by Hylleraas, the angular quantities were replaced by only metric ones by means of the formulae:
\be
a_{1} = \frac{r_1^2 + r_{12}^2 - r_2^2}{2 r_1 r_{12}}, \quad
a_{2} = \frac{r_2^2 + r_{12}^2 - r_1^2}{2 r_2 r_{12}}. 
\ee
For some reason, apparently Majorana did not completed the full calculations (see \cite{quadtep4}), which would have resulted in the following expression for the energy to be minimized:
\be \label{compli}
W = 
\frac{2 (k-\ell) \left[8 k^3+k^2 (5 -7 \ell-16 Z)+4 k \ell (Z + \ell -1)- \ell^2 (\ell-1) \right]}{8 k^2-5 k \ell+\ell^2} .
\ee
As expected, the minimization procedure gives a better result\footnote{$-W/{\rm Rh}=5.779$ instead of $-W/{\rm Rh}=729/128=5.695$.} for the helium ground state energy with respect to the $\ell=0$ case. However, despite the apparent complicated formula in (\ref{compli}), the striking fact is the accuracy of the Majorana result, obtained just at {\it first order}, $W_I^{\rm He} =24.09$ eV (corresponding to $k=1.8581$, $\ell=0.2547$), when compared with the analogous one produced by Kellner at {\it fourth order}, $W_I^{\rm He} =23.75$ eV, and even with that obtained by Hylleraas with calculations at {\it order 11}, $W_I^{\rm He} =24.35$ eV.

\section{Conclusions}
\label{s4}

\noindent The early history of quantum mechanics was indissolubly related to the successful description of the helium atom, with particular reference to the prediction of its ground state energy. The basic difference with the description of the simplest atomic system, the hydrogen (or hydrogenoid) atom, is that the two-electron problem does not admit an analytic solution, so that quantitative predictions about it are {\it de facto} mediated by the approximation method employed to get numerical estimates, to be then compared with the experimental observations. 

As seen in Sect. \ref{s2}, though depending on the chosen unperturbed system, perturbative calculations performed since 1926 did not give very accurate predictions, even by increasing the perturbative order, the reason for this (envisaged by Slater) being due to the form of the Hamiltonian operator for the system at hand. Such an occurrence led physicists to adopt different approaches, the first of which was the variational method early introduced by Ritz. Calculations performed by Kellner in 1927, indeed, did improve perturbative results, but the interesting novel idea introduced by him was that of an effective nuclear charge experienced by one of the two electrons in the field of the nucleus plus the other electron, which shields to some extent the nuclear charge. 

In the meanwhile, at the end of 1927 Hartree elaborated his self-consistent field method aimed at describing many-eletron atoms, whose basic assumption was just that such atoms could be described by wavefunctions given as the product of one-electron wavefunctions of individual electrons in the field of a corrected (``self-consistent") central potential. Although criticized by many, mainly due to its ``unusual" line of reasoning, the Hartree method gave a prediction for the helium ionization potential very close to the experimental determinations.

Important novelties on the subject came early in 1928 by means of Hylleraas, who realized that the absolute orientation of the atom in the space was irrelevant, this leading to an enormous yet unexpected simplification into the variational problem, described only in terms of three variables defining the shape of the electron-nucleus triangle, instead of the six spatial coordinates of the two electrons. With this simplification, Hylleraas was able to perform order 11 calculations (instead of only order 4 ones by Kellner), which provided an excellent approximation for the ionization energy of helium. The method and reasoning elaborated by Hylleraas, with the subsequent introduction of what will become known as Mandelstam variables $s,t,u$, were worth to be easily generalized, and the later introduction of screening constants further improved the predictions. The quantum mechanical two-electron problem converted, around 1930, just into a numerical one, whose accuracy depended only on the degree of approximation used.

This known history of the solutions given to the two-electron problem has been here supplemented by several, unpublished results contributed by the Italian physicist Ettore Majorana around the same years, as described in detail mainly in Sect. \ref{s3}. Indeed, almost all of these results did not convey (even indirectly) in his two papers on the topics, published in 1931, so that they remained unknown until now. Due to their simple deduction, Majorana's reproduction of standard perturbative and variational results, but generalized to arbitrary $Z$, has been reported as well (in Sect. \ref{s2}) for pedagogical reasons: they are summarized by the formulae in Eqs. (\ref{first}) and (\ref{EMvar}), respectively. 

From a purely theoretical point of view, the most intriguing contribution is a variant of the variational method,  developed by Majorana in order to take directly into account, already in the trial wavefunction, the action of the full Hamiltonian operator of the given system. From the general wavefunction in Eq. (\ref{EMgenvar1}) he was, then, able to obtain a determinantal equation (\ref{EMgenvar2}) from which the ground state energy could be deduced. Though this method was originally devised for helium (or, more generally, for the two-electron problem), it is well suited for application to {\it any} variational problem and, as such, it is worth to be applied in different fields of present day research.

Instead, coming back to the problem of the ground state energy of two-electron atoms, it is particularly interesting to mention an application of the perturbative method where the atomic number $Z$ was treated effectively as a {\it continuous} variable, contributions to the ground state energy of an atom with given $Z=Z_0$ coming also from any {\it other} $Z \neq Z_0$. Its application to several empirical relations aimed at finding accurate predictions - without the use of complicated unperturbed wavefunctions - has been described in detail in Sect. \ref{3.2}.

The use of  particular wavefunctions, along a line of reasoning similar to that adopted by Slater in 1928, was as well considered by Majorana. In addition to the conclusions achieved by Slater, regarding the accuracy of the perturbative method (obtained, however, in a more transparent and quantifiable manner, with the deduction of broad range estimates), he introduced the concept of an effective nuclear charge {\it different} for the two electrons, thus generalizing previous results (see Sect. \ref{3.3}).

Finally, we have discussed a simple alternative to Hylleraas' method which, as explained in Sect. \ref{3.4}, led Majorana to obtain first order numerical results comparable in accuracy with those by Hylleraas at order 11 (and certainly better results with respect to those by Kellner at fourth order), just by choosing a suitable (and quite ``obvious") trial wavefunction. 

Summing up, if the role played by Majorana's contributions in the known literature on the two-electron problem was, in a sense, certainly subdominant, nevertheless his unpublished results add new non negligible pieces to the history of early quantum mechanics. But, especially, their pedagogical relevance in undergraduate and postgraduate courses, as well as the importance of novel methods in current researches in quantum physics, as described in the present paper, is yet to be fully discovered and exploited.


\end{document}